\documentclass[a4paper,twoside]{article}
\newcommand{\affil}[1]{$^{\rm #1}$}
\usepackage[authoryear]{natbib}
\bibpunct{(}{)}{;}{a}{}{,}
\usepackage{graphicx}
\date{} %Please leave the date blank
\title{\large\bf\flushleft Pulsars and Gravitational Wave Detection}
\author{\parbox{\textwidth}{\flushleft
\vspace{-0.5cm}
{\it George Hobbs\affil{A}}\\
\vspace{0.4cm}
{\small \affil{A}\,Australia Telescope National Facility, CSIRO, PO
  Box 76, Epping, NSW 1710, Australia}\\
{\small \affil{A}\,Email: george.hobbs@csiro.au}}}
\begin{document}
\twocolumn[
\begin{changemargin}{.8cm}{.5cm}
\begin{minipage}{.9\textwidth}
\vspace{-1cm}
\maketitle
%
%
%%%%%%%%%%%%%     ABSTRACT    %%%%%%%%%%%%%
%Abstract of no more than 200 words here.
\small{\bf Abstract:} 

The number of known millisecond pulsars has dramatically increased in
the last few years.  Regular observations of these pulsars may allow
gravitational waves with frequencies $\sim 10^{-9}$\,Hz to be
detected.  A ``pulsar timing array'' is therefore complimentary to
other searches for gravitational waves using ground-based or
space-based interferometers that are sensitive to much higher
frequencies.  In this review we describe 1) the basic methods for
using an array of pulsars as a gravitational wave detector, 2) the
sources of the potentially detectable waves, 3) current limits on
individual sources and a stochastic background and 4) the new project
recently started using the Parkes radio telescope.

%%%%%%%%%%%%%     KEYWORDS    %%%%%%%%%%%%%
\medskip{\bf Keywords:} pulsars: general --- gravitational waves
% Please write all keywords in lower case. PASA uses the
% standard list of subject headings adopted by The Astrophysical Journal
% and available from http://www.journals.uchicago.edu/ApJ/keywords_text.html.
% Keywords are separated by en-dashes, i.e. --

%%%%%%%%DO NOT EDIT%%%%%%%%%%%%
\medskip
\medskip
\end{minipage}
\end{changemargin}
]
\small
%%%%%%%%EDIT FROM HERE%%%%%%%%%%%%

\section{Introduction}
%Please see the PASA Style Guide for help with correct layout for your manuscript.
%Examples of tables and figures are given below.

\begin{figure}[h]
\begin{center}
\includegraphics[scale=0.25, angle=-90]{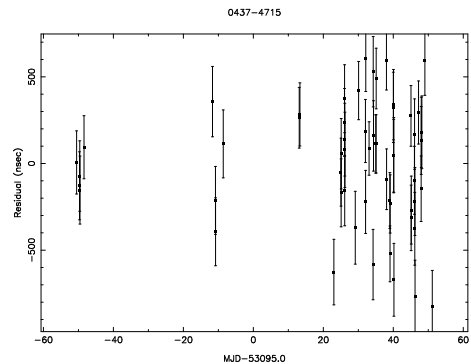}
\caption{Timing residuals (the difference between actual pulse arrival
  times and times predicted using a spin-down model for the pulsar)
  for PSR~J0437$-$4715. The timing residual root-mean-square (rms) is
  400\,ns.}\label{fg:0437}
\end{center}
\end{figure}

The assumption that a pulsar is a regular rotator that follows a
predictable slow-down model forms the basis of a powerful technique
for finding its rotational and positional properties.  This ``pulsar
timing'' technique (see e.g. Manchester \& Taylor 1977\nocite{mt77}
for a general review or Blandford, Narayan \& Romani
1984\nocite{bnr84} and Backer \& Hellings 1986\nocite{bh86} for more
details) allows the arrival times of pulses from a particular pulsar
to be predicted with great accuracy.  For some pulsars that have spin
periods of a few milliseconds (hereafter referred to as the
``millisecond pulsars'' or MSPs) the pulse arrival times can be
modelled to less than a microsecond over many years of observations
(see Figure~\ref{fg:0437}).  Due to this phenomenal precision,
gravitational waves (GWs) with periods between days and decades should
be detectable by analysing slight discrepancies between predicted and
actual pulse arrival times. As emphasised by Foster
(1990)\nocite{fos90}, the effect of a GW passing a free test mass,
such as the Earth or a pulsar, is not to move the mass from its
coordinate position, but instead to deform the space-time metric
around the mass.
 
% As summarised in Kitchen (1998) and many texts on general relativity,
% the basic concept of a GW is simple.  The gravitational influence of a
% massive object will vary as the body moves and information on that
% changing gravitational field will propagate outward through the
% space-time continuum at the speed of light in a manner that obeys
% equations analogous to those for electromagnetic radiation.  Although
% the general theory of relativity and other post-Newtonian
% gravitational theories all predict the existence of GWs, details
% differ between the theories. For instance, Einstein's theory forbids
% dipole radiation, but this is allowed by many of the other
% theories. As emphasised by Foster (1990)\nocite{fos90}, the effect of
% a GW passing a free test mass, such as the Earth or a pulsar, is not
% to move the mass from its coordinate position, but instead to deform
% the space-time metric around the mass.

Pulsar studies have already placed stringent limits on a GW background
and have been used to rule out, or place limits on, some cosmic string
models. As will be shown, pulsar studies have also placed constraints
on postulated supermassive black hole binary systems in our and nearby
galaxies. In section 2, we describe the basic framework for studying
the observable effect of GWs on pulsar timing residuals.  Section 3
contains a discussion on the creation and detection of a stochastic
background of GWs.  In section 4, we describe how individual sources
of GWs may be detected. In section 5 we highlight some practical
issues neccessary for the detection of GWs. We conclude with a
description of the current status of the Australian timing array
project.

\section{Basics}

 Detweiler (1979)\nocite{det79} provided the basic framework for
 describing the effect of a GW passing through the solar system on a
 pulsar's timing residuals.  In brief, the pulsar and Earth should be
 considered as the ends of a free-mass GW antenna.  In order to detect
 GWs the relative motion of the pulsar and Earth must be monitored by
 observing fluctuations in the pulsar's observed spin rate. The
 measured frequency $\nu(t)$ of a pulsar of constant frequency
 $\nu_0$, with direction cosines $\alpha$, $\beta$ and $\gamma$ varies
 slightly as a GW passes the solar system as (Detweiler 1979)
 \begin{eqnarray}
  \label{eqn:det1}
  z(t) &=& \frac{\nu_0 - \nu(t)}{\nu_0}  \\ \nonumber
       &=& \frac{(\alpha^2 -
  \beta^2)(h_+^E - h_+^P) + 2\alpha\beta (h_X^E -
       h_X^P)}{2(1+\gamma)} \\ \nonumber
 \end{eqnarray}
 where $h_+^{E,P}$ and $h_X^{E,P}$ are the wave amplitudes in the two
 polarizations at the Earth (E) and pulsar (P). 

 The observable effect of a GW is therefore to create pulse period
 fluctuations with an amplitude proportional to the gravitational wave
 strain evaluated at the Earth, $h_{+,X}^E$, and at the pulsar
 $h_{+,X}^P$.  However, the values of $h_{+,X}^P$ for widely-spaced
 pulsars will be uncorrelated, whereas the component at the Earth will
 be correlated.  It is therefore possible to obtain this correlated
 signal by combining measurements from multiple pulsars; i.e. by using
 observations from a pulsar timing array.  Hellings \& Downs (1983)
 developed a simple method for determining the GW signal common to all
 pulsars by cross-correlating the time derivative of the timing
 residuals for multiple pulsars.  We note that the Doppler shifts in
 the apparent rotational rates of pulsars are correlated around the
 sky with a quadrupole and higher order angular signature. More
 details about combining multiple data sets to search for this
 signature have been described by Foster (1990).

 \section{Stochastic Backgrounds} 

 \subsection{Creation}

  A stochastic background of GWs can be cosmological (e.g. due to
  inflation, cosmic strings or phase transitions), or astrophysical
  (e.g. due to coalescing massive black hole binary systems that result
  from the mergers of their host galaxies).

  Models for cosmological stochastic backgrounds of GWs have been
  reviewed in Maggiore (2000)\nocite{mag00}.  Such GWs can have
  frequencies between $f \sim 10^{-18}$\,Hz (corresponding to a
  wavelength as large as the present Hubble radius of the Universe) to
  $f \sim 10^{+12}$\,Hz (which corresponds to the frequency of a
  graviton produced during the Planck era and redshifted to the present
  time using the standard cosmological model).  One mechanism for
  producing copious amounts of GWs is based upon topological defects
  that formed during phase transitions in the early universe (``cosmic
  strings'').  The predicted GW spectrum due to these cosmic strings
  has an almost flat region that extends from $f \sim 10^{-8}$\,Hz to
  $10^{10}$\,Hz and a peak in the region of $f \sim 10^{-12}$\,Hz.
  Details of the implications of the current pulsar timing limit are
  provided by Caldwell, Battye \& Shellard (1996)\nocite{cbs96} who use
  the limit to exclude a range of values for the cosmic string linear
  mass density for certain values of cosmic string and cosmological
  parameters.

  \begin{figure}[h]
  \begin{center}
  \includegraphics[scale=0.4]{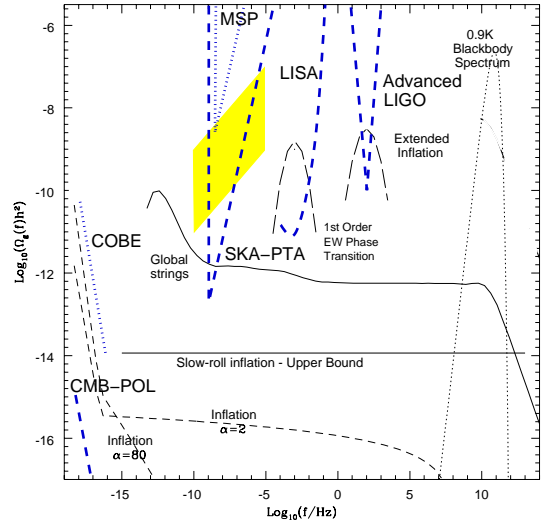} 
  \caption{Predicted gravitational wave backgrounds.  The top of the figure
  contains current and predicted limits placed on the gravitational
  wave background from millisecond pulsar timing, LISA and advanced LIGO.
  Figure provided by M. Kramer.}\label{fg:kramer}
  \end{center}
  \end{figure}
 
 Stringent limits have already been placed on the energy density of a
 stochastic background using the timing residuals of individual
 pulsars.  For a flat GW energy spectrum that is centred on some
 frequency $f$ and has a bandwidth also equal to $f$ then an upper
 limit on the energy density ($\rho_g$) of a GW background can be
 obtained from (Detweiler 1979)

 \begin{equation}
  \rho_g < \frac{243}{208}\frac{\pi^3f^4}{G}\langle R^2(t)\rangle
  \label{eqn:det2}
 \end{equation}
 where $G$ is Newton's gravitational constant and $\langle R^2(t)
 \rangle$ the rms timing residual. Romani \& Taylor
 (1983)\nocite{rt83} obtained an upper limits on the equivalent mass
 density in GWs in frequency ranges between $0.7 \times 10^{-8}$ and
 $6 \times 10^{-8}$\,Hz and concluded that, at the present epoch, the
 mass density of the universe is not dominated by GWs of frequency
 $\sim 10^{-8}$\,Hz. 

 The expected power spectrum for cosmological models was shown by
 Blandford et al. (1984) to be

 \begin{equation}
  P_g(f) = \frac{G\rho_g(f)}{3\pi^3f^4} = \frac{H_0^2}{8 \pi^4}\Omega_gf^{-5}  
  \label{eqn:kaspi1}
 \end{equation}
 in which $\rho_g(f)$ is the energy density of the stochastic
 background at frequency $f$, $H_0 = 100h_0$\,km\,s$^{-1}$\,Mpc$^{-1}$
 is the Hubble constant and $\Omega_g$ the fractional energy density
 in GWs per logarithmic frequency interval.  Therefore, if $\Omega_g$
 is constant then $P(f) \propto f^{-5}$.  The index of this power-law
 ($\alpha = 5$) should be contrasted with $\alpha = 0$ for white noise
 and $\alpha = 2 \rightarrow 4$ expected for clock instabilities,
 ephemeris errors, interstellar propagation effects and pulsar
 rotational instabilities (see Stinebring et al. 1990). Stinebring et
 al. (1990) used seven years of observations to place rigorous upper
 bounds on the stochastic background. Using a similar method and seven
 years of data for PSR~B1855+09 allowed Kaspi, Taylor \& Ryba (1994)
 to place a limit of $\Omega_{gw}h_0^2 < 6 \times 10^{-8}$.  McHugh et
 al. (1996) provided a more statistically sound method to obtain that
 $\Omega_{gw}h_0^2 < 10^{-6}$.  Their result is independent of the
 assumption of a flat spectrum for $\Omega_{gw}(f)$.

 An astrophysical background would be formed by GW radiation from
 supermassive black holes. Current theories suggest that galaxies
 contain a central black hole of mass $\stackrel{>}{_\sim}
 10^6$\,M$_\odot$.  As many galaxies are observed to be merging, the
 existence of a binary black hole system in a merger remnant is
 likely.  If the binary loses enough energy and angular momentum then
 it may enter a regime where gravitational radiation alone can bring
 about inspiral and coalescence (Rajagopal \& Romani
 1995)\nocite{rr95}.  During an inspiral the GWs will sweep through a
 range of frequencies.  The detection of such events clearly depends
 upon the rate of occurrence and the amplitude and frequencies of the
 GWs produced.  The possibility of detecting a stochastic background
 of such events with pulsar timing was described by Rajagopal \&
 Romani (1995).

\begin{figure}[h]
 \begin{center}
 \includegraphics[scale=0.3]{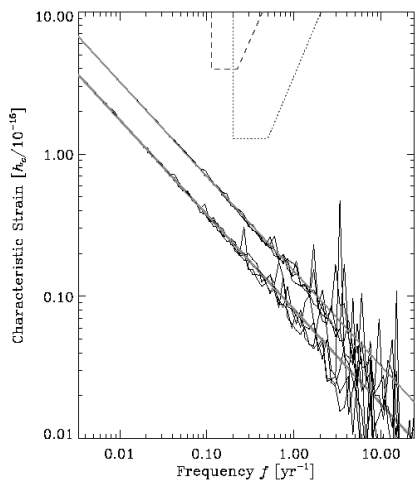} 
 \caption{Characteristic strain spectrum for different models of a GW
 background formed due to merging black-hole binary systems. The
 dashed line gives the current best limits on the background from
 pulsar timing observations and the dotted line provides the expected
 limits from the pulsar timing array after $\sim$\,8 years of
 operation. Figure obtained from Jaffe \& Backer
 2003.}\label{fg:j_backer}
 \end{center}
 \end{figure}

 This work was continued by Jaffe \& Backer (2003)\nocite{jb03} who
 find that the spectrum of a stochastic background of black hole
 binary systems has a characteristic strain of $h_c(f) \sim
 10^{-16}(f/yr^{-1})^{-2/3}$ which is just below the detection limit
 from recent analyses of pulsar timing measurements (see
 Figure~\ref{fg:j_backer})\footnote{This characteristic strain
 $h_c(f)$ is defined as $h_c(f) = \sqrt{fS_h(f)}$ where $S_h$ is the
 spectral density with units of inverse frequency.  This can be
 related to $\Omega_{g}(f)$ by
 \begin{equation}
  \Omega_{g}(f) = \frac{2\pi^2}{3H_0^2}f^3S_h(f)
 \end{equation}}.  Even though the
 amplitude of this spectrum was considered too high by Wyithe \& Loeb
 (2003)\nocite{wl03}, this background is likely to dominate over
 cosmological stochastic backgrounds and is therefore the most likely
 background to be detected using a pulsar timing array.  The actual
 background reached can be determined from the slope of the spectrum
 which will indicate whether we are observing a coalescing population
 or other sources of stochastic GWs.

 \section{Individual sources}

 A periodic source of gravitational radiation will produce a periodic
 shift in the pulse arrival time.  Sufficient GW amplitudes and
 frequencies that can potentially be detected using a pulsar timing
 array are predicted to occur from a supermassive black hole binary
 system in a nearby galaxy (or in the centre of our Galaxy).  The
 expected signature for a supermassive black hole binary system with a
 circular orbit is given by (Jaffe \& Backer 2003)

 \begin{equation}
  h \approx 4.4\times10^{-17}m_8^{5/3}P_{yr}^{-2/3}d_{Gpc}^{-1}\frac{q}{(1+q)^2}
 \end{equation} 
 where $h$ is an order-of-magnitude estimate of the strain amplitudes,
 $m_8$ is the total mass of the system in units of $10^8$M$_\odot$,
 $P_{yr}$ the observed GW period in years, $d$ the distance to the
 emitter from the Earth in GPc and $q$ is the mass ratio ($q < 1$).
 They also find that the power radiated along the axis of the orbit is
 eight times that for an edge-on view.  The actual effect on the
 pulsar arrival times will depend upon the angle between the pulsar GW
 source and the pulsar; a pulsar lying along the line of sight to the
 GW source will experience no effect.

 Lommen \& Backer (2001)\nocite{lb01} searched for gravitational
 radiation from Sagittarius A$^*$ which had been postulated to be a
 massive black-hole binary system (see e.g. Zhao, Bower \& Goss
 2001\nocite{zbg01}).  They calculated that the expected effect would
 be about $\sim$\,10ns in the timing residuals of PSRs~B1937$+$21 and
 J1713$+$0747 which is too small to be detectable with current data.
 Lommen \& Backer (2001) tabulated the expected timing residuals for
 postulated binary massive black holes in nearby galaxies assuming an
 equal-mass binary system with an orbital period of 2000\,days.  If we
 can identify structures with amplitudes of $\sim$100\,ns in the
 timing residuals then meaningful constraints can be placed on
 $\sim$10 nearby sources.

\begin{figure}[h]
 \begin{center}
 \includegraphics[scale=0.4]{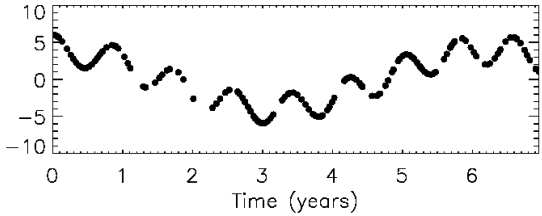} 
 \caption{Theoretical timing residuals induced by GWs from 3C66B. No
 such signature was observed in real pulsar timing residuals (Figure
 obtained from Jenet et al. 2004)}\label{fg:jenet}
 \end{center}
 \end{figure}

 The expected signature for the timing residuals for the more general
 case of a coalescing, binary system in an eccentric orbit was
 presented in Jenet et al. (2003)\nocite{jllw04}. The effect depends
 upon the orbital parameters (including the orbital inclination
 angle), masses, source distance and the opening angle between the
 source and the pulsar relative to Earth.  Jenet et al. (2004)
 attempted to detect GWs emitted by the proposed supermassive binary
 black hole system in 3C66B (Sudou et al. 2003\nocite{simt03}). The
 expected signature in the timing residuals of PSR~B1855+09 are two
 sinusoids, one with an amplitude $\sim 5\mu s$ and a period of
 0.88\,years and the other of amplitude $\sim 10\mu s$ with a
 6.2\,year period (see Figure~\ref{fg:jenet}).  The two sinusoids
 occur if the Earth-pulsar line-of-sight is perpendicular to the GW
 propagation vector as the observed timing residuals will contain
 information about the source at the current epoch and 4000 years ago
 (the distance to PSR~B1855+09 from Earth is 4000\,lt-yr).  No such
 signature was found.

 \section{Practical issues}

 Pulsar timing is affected by the stability of terrestrial clocks,
 ephemeris errors and the pulsar itself.  In order to make a
 definitive detection of gravity waves then a timing array requires a
 minimum of five pulsars spread over the sky (see Foster \& Backer
 1990). One pulsar is required to confirm the stability of terrestrial
 clocks, three widely spaced pulsars are necessary to solve for an
 error in the solar system ephemeris and the fifth pulsar to place an
 upper limit on the GW background.  To solve completely for all the
 available information about the background, more pulsars are
 required.

 The pulse arrival times must be determined to high precision (current
 timing array projects are aiming to achieve a precision between
 $10^{-9}$ and $10^{-6}$\,seconds). The necessary precision can be
 estimated from (Rajagopal \& Romani 1995)\nocite{rr95}

 \begin{equation}
  h \sim \frac{\delta t}{P_w}(N_{obs}T_{obs})^{-1/2}
 \end{equation}
 in which $h$ is the typical strain amplitude of a detectable GW,
 $N_{obs}$ the number of arrival times measured each year, $T_{obs}$
 the total length of time that the pulsar has been observed, $P_w$ the
 period of the GW and $\delta t$ the typical arrival time
 precision. Such timing is achievable. For example, van Straten et
 al. (2001)\nocite{vbb+01} obtained a residual root-mean-square of
 only 130\,ns over 40\,months of observing PSR~J0437$-$4715. With
 better instrumentation this precision should be improved. Measuring
 pulse arrival times with high precision for most MSPs benefits from
 observing at frequencies of $\sim 3$\,GHz or higher to counteract
 interstellar propagation effects (Rickett 1977\nocite{ric77}). It is
 also essential that the pulsar's dispersion measure is known
 accurately for every observation. To do this, simultaneous multiple
 frequency observations are required at widely spaced observing
 frequencies.

 The pulsars chosen as part of a timing array must be intrinsically
 stable. Some pulsars show ``timing noise'', a continuous, noise like
 fluctuation in the rotation rate (for example, Lyne 1999\nocite{lyn99})
 or glitches, sudden increases in rotation rate (Lyne, Shemar \&
 Graham-Smith 2000\nocite{lsg00}). However, the MSPs have been shown
 to be extremely stable over many years of observing. In fact, the
 stability of many MSPs rival that of terrestrial atomic time
 standards (for example, see Lommen 2002\nocite{lom02}).

 The power spectrum of the time derivative of the pulsar timing
 residuals must be obtained in order to obtain detailed information
 about the stochastic background.  Methods for obtaining this power
 spectrum from the irregularly sampled pulsar timing residuals have
 been fiercely debated in the literature.  Stinebring et
 al. (1990)\nocite{srtr90} developed a method using orthonormal
 polynomials as basis functions.  This work was generalised for
 multiple pulsar data sets by Kaspi, Taylor \& Ryba
 (1994)\nocite{ktr94}, but was criticised by Thorsett \& Dewey
 (1996)\nocite{td96} who developed a technique based on the
 Neyman-Pearson test.  McHugh et al. (1996)\nocite{mzvl96}
 subsequently showed the Neyman-Pearson test of hypothesis also
 cannot, in the general case, provide upper limits on an unknown
 parameter and suggested the use of a Bayesian formalism.

 When forming the power spectrum of the pulsar timing residuals it is
 important to recall that the residuals were obtained by fitting a
 model that includes at least the phase, spin-period, its derivative
 and position to the pulse arrival times.  For many MSPs, fits have
 also been made for the system's orbital parameters.  The fitting
 procedure therefore removes long-period variations in the arrival
 times and hence limits on GWs are not valid for GW periods near to
 the length of the data span (see Backer \& Hellings
 1986)\nocite{bh86}.  The transfer function of the pulsar model as a
 filter was obtained by Blandford, Narayan \& Romani (1984)\nocite{bnr84} and
 should therefore be taken into account when studying relatively short
 data spans.

\section{The Australian pulsar timing array}

\begin{figure}[h]
\begin{center}
\includegraphics[scale=0.25, angle=-90]{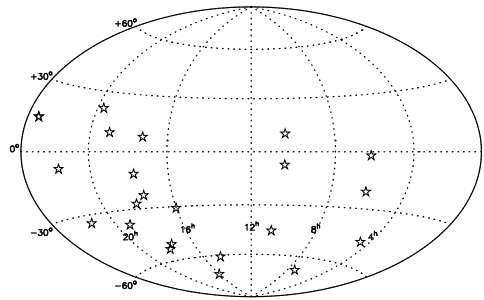}
\caption{Location of the millisecond pulsars on the sky that are being
observed as part of the Australian pulsar timing array project.}\label{figexample}
\end{center}
\end{figure}

Since Februrary 2004, in a collaborative effort between the ATNF and
the Swinburne and Caltech universities, we have been observing
$\sim$20 millisecond pulsars using the new 680/3100\,MHz
dual-frequency receiver and a 1400\,MHz receiver at the 64-m Parkes
radio telescope (see http://www.atnf.csiro.au/research/pulsar/array).
We intend to observe each pulsar at 7--10 day intervals over a period
of at least five years.  For many of these pulsars we already have
timing residual rms values less than 1$\mu s$ with observation times
of 30 minutes or 1 hour depending upon the brightness of the
pulsar. For PSR~J0437$-$4715 we currently have 166 dual-frequency
observations spanning 125 days which give us, with only rudimentary
processing being applied to the data, a timing residual rms of
$\sim$500\,ns.  This short data span already places a limit of
$\Omega_{gw}h_0^2 < 5\times10^{-4}$ on any possible existence of a GW
background.  If we can reduce the timing residual to 100\,ns over 5
years then the limit (from a single pulsar) will be $\Omega_{gw}h_0^2
< 5\times10^{-10}$ and will provide tight constraints on gravitational
wave backgrounds from merging black holes and cosmic strings.

We have also been developing a software package, \textsc{superTempo},
for processing the arrival times from multiple pulsars
simultaneously. The current status of this, and other related
projects, may be found on our web-site.

\section{Conclusion}

 It is likely that gravitational waves will be detected within the
 next few years by world-wide pulsar timing array projects.  As a
 by-product of these investigations stringent checks will also be
 placed on terrestrial time standards and the solar system
 ephemeris. The regular dual-frequency observations of multiple
 pulsars will also provide valuable information about the
 interstellar medium. Using a pulsar array as a gravitational wave
 detector is complimentary to other searches currently being designed
 that are attempting to detect much shorter-period GWs.   

%%Format tables as in the following example
%\begin{table}[h]
%\begin{center}
%\caption{Example Table}\label{tableexample}
%\begin{tabular}{lcc}
%\hline Column 1 & Column 2 & Column 3 \\
%\hline Table Content$^a$ \\
%\hline
%\end{tabular}
%\medskip\\
%$^a$Table footnotes go here.\\
%\end{center}
%\end{table}

\section*{Acknowledgments} %If needed

 I thank R. Manchester, F. Jenet and M. Kramer for providing useful
 comments and suggestions.

%\bibliography{modrefs,psrrefs}
%\bibliographystyle{mn}

% kitchen, 
%\end{multicols}

\end{document}